\documentclass[preprint2]{aastex}	
\usepackage{psfig}

\def\gradip{\hbox{\rlap{\hbox{.}}\raise 5.truept \hbox{{\small $\circ$}}}}
\def\pn{\par\noindent}

\catcode`\@=11
\def\gsim{\ifmmode{\mathrel{\mathpalette\@versim>}}
    \else{$\mathrel{\mathpalette\@versim>}$}\fi}
\def\lsim{\ifmmode{\mathrel{\mathpalette\@versim<}}
    \else{$\mathrel{\mathpalette\@versim<}$}\fi}
\def\@versim#1#2{\lower 2.9truept \vbox{\baselineskip 0pt \lineskip
    0.5truept \ialign{$\m@th#1\hfil##\hfil$\crcr#2\crcr\sim\crcr}}}
\catcode`\@=12

\begin{document}

\title{The Proper Motion of the Globular Cluster NGC~6553 and of Bulge
Stars with HST\altaffilmark{1}}

\author{
M. Zoccali\altaffilmark{2}, A. Renzini\altaffilmark{2},
S. Ortolani\altaffilmark{3}, E. Bica\altaffilmark{4},
B. Barbuy\altaffilmark{5}
}

\altaffiltext{1}{This work is based on observations with the NASA/ESA Hubble
Space Telescope, obtained at the Space Telescope Science Institute,
operated by AURA Inc. under contract to NASA}

\altaffiltext{2}{European Southern Observatory, Karl Schwarzschild Strasse 2,
D-85748 Garching bei M\"unchen, Germany; mzoccali@eso.org, arenzini@eso.org}

\altaffiltext{3}{Universit\`a di Padova, Dipartimento di Astronomia, Vicolo
dell'Osservatorio 5, I-35122 Padova, Italy; ortolani@pd.astro.it}

\altaffiltext{4}{Universidade Federal do Rio Grande do Sul, Dept. de Astronomia,
CP 15051, Porto Alegre 91501-970, Brazil; bica@if.ufrgs.br}

\altaffiltext{5}{Universidade de S\~ao Paulo, Dept. de Astronomia,  CP 3386,
S\~ao Paulo 01060-970, Brazil; barbuy@orion.iagusp.usp.br}

\begin{abstract}

WFPC2 images obtained with the Hubble Space telescope 4.16 years apart
have allowed us to measure the proper motion of the metal rich
globular cluster NGC~6553 with respect to the background bulge stars.
With a space velocity of (${\Pi},~{\Theta},~W$) = (-3.5, 230, -3) km
s$^{-1}$, NGC~6553 follows the mean rotation of both disk and bulge
stars at a Galactocentric distance of 2.7 kpc.  While the kinematics
of the cluster is consistent with either a bulge or a disk membership,
the virtual identity of its stellar population with that of the bulge
cluster NGC~6528 makes its bulge membership more likely.  The
astrometric accuracy is high enough for providing a measure of the
bulge proper motion dispersion and confirming its rotation. A
selection of stars based on the proper motions produced an extremely
well defined cluster color-magnitude diagram (CMD), essencially free
of bulge stars.  The improved turnoff definition in the decontaminated
CMD confirms an old age for the cluster ($\sim 13$ Gyr) indicating
that the bulge underwent a rapid chemical enrichment while being built
up at in the early Universe. An additional interesting feature of the
cluster color-magnitude diagram is a significant number of blue
stragglers stars, whose membership in the cluster is firmly
established from their proper motions.

\end{abstract}


\section{INTRODUCTION}

The globular cluster system of the Milky Way has a slightly bimodal
metallicity distribution, and it was been suggested that the metal
rich group ([Fe/H]$>-0.8$) would constitute a population of ``disk''
globular clusters (Zinn 1985; Armandroff 1989). These clusters are
actually strongly concentrated towards the Galactic center, with most
of them being physically inside the Galactic bulge, at a
galactocentric distance $\lsim 3$ kpc (see e.g., Fig. 3 in Zinn 1985).
Much progress has been made in recent years in the study of this group
of clusters, thanks to higher angular resolution CCD and near-infrared
photometry and high resolution spectroscopy. The kinematics, vertical
scale height, and metallicity distribution of these clusters turn out
to be indistinguishable from those of the bulge field stars (Minniti
1995; Barbuy et al. 1998, 1999b; C\^ot\'e 1999), leaving little doubt
that they constitute a family of bulge, rather than disk globular
clusters (see also Harris 2000 for a recent discussion).

Among these clusters, the best studied ones are the ``twin'' clusters
NGC~6553 and NGC~6528 which have virtually identical color-magnitude
diagrams (Ortolani et al. 1995).  The subject of the present
investigation, NGC~6553 ($\alpha$=18:09:15.6, $\delta$=-25:54:28,
$l=5\gradip25$, $b=-3\gradip02$), is a moderately concentrated
(central surface brightness: log$\Sigma =4.53 L_\odot pc^{-2}$;
Djorgovski 1993) globular cluster located at $\sim 3$ kpc from the
Galactic Center. Its metallicity is high, although the exact value is
still a matter of debate: Barbuy et al. (1999a) and Coelho et al
(1999) found [Fe/H]=$-0.5$, while Cohen et al. (1999) measured
[Fe/H]=$-0.16$. The first CCD photometry in the $BVRI$ bands showed
clear metal-rich characteristics, with the pronounced $V$ band
luminosity turn-over of the upper Red Giant Branch (RGB) towards
cooler temperatures, an effect of strong TiO blanketing (Ortolani et
al. 1990).  Subsequent $VI$ data obtained with the Hubble Space
Telescope (HST) by Ortolani et al.\ (1995; see also Guarnieri, Renzini
\& Ortolani 1997; Guarnieri et al. 1998) reached more than 3
magnitudes below the turnoff, leading to the conclusion that NGC~6553
is nearly coeval to the halo globular clusters.  Still, the very high
contamination by background stars around the cluster turnoff was very
high, leaving some uncertainty on the value of the turnoff luminosity
$M_V^{TO}$.  The more recent study by Sagar et al.\ (1999) over a
larger field ($6^\prime \times 10^\prime$) allowed them to obtain a
cleaner cluster population by statistical subtraction of background
stars. However, their data are not deep enough to allow photometry of
turnoff stars with the accuracy required to address the problem of the
cluster age.

Using a technique pioneered by King et al.\ (1998), in the present
work we combine HST wide field and planetary camera (WFPC2) data from
two epoch observations of the central regions of NGC~6553.  Thanks to
the excellent astrometric performances of WFPC2, we measure the
relative proper motion of the cluster with respect to the bulge. This
allows us to {\it i}) decontaminate the CMD of NGC~6553 from the
background and therefore to obtain a more reliable measure of its age;
{\it ii}) obtain the three components of the cluster space velocity;
and {\it iii}) measure the bulge proper motion dispersions along
Galactic longitude and latitude.

\section{OBSERVATIONAL DATA AND ANALYSIS}

\begin{table}
\begin{center}
\tablewidth{0pt}
\begin{tabular}{cccr}
\multicolumn{4}{c}{\small TABLE ~ 1} \\
\multicolumn{4}{c}{\small LOG OF OBSERVATIONS} \\
\hline\hline
Program & Date & Filter & Exp. (s) \\
\hline
GO5436 & 25.2.1994 & F555W & 14           \\
       &           & F555W & 2$\times$100 \\
       &           & F814W & 5            \\
       &           & F814W & 2$\times$50  \\
GO7307 & 24.4.1998 & F555W & 3$\times$5   \\
       &           & F555W & 3$\times$200 \\
       &           & F814W & 3$\times$20  \\
       &           & F814W & 3$\times$200 \\
\hline\hline
\end{tabular}
\end{center}
\end{table}

We analyse two sets of observations of the central region of the bulge
globular cluster NGC~6553, taken with a time interval of 4.16 years.
Both sets were collected with HST using the WFPC2 camera, with the
filters F555W ($V$) and F814W ($I$).  The first epoch data were
acquired by our group on February 1994, as part of the GO5436 program
(Ortolani et al. 1995; Guarnieri et al. 1997).  Those of
the second epoch were collected in April 1998, as part of the GO7307
program, and were retrieved from the HST archive.  The log of
observations for the two epochs is shown in Table~1.

\begin{figure*}
\psfig{file=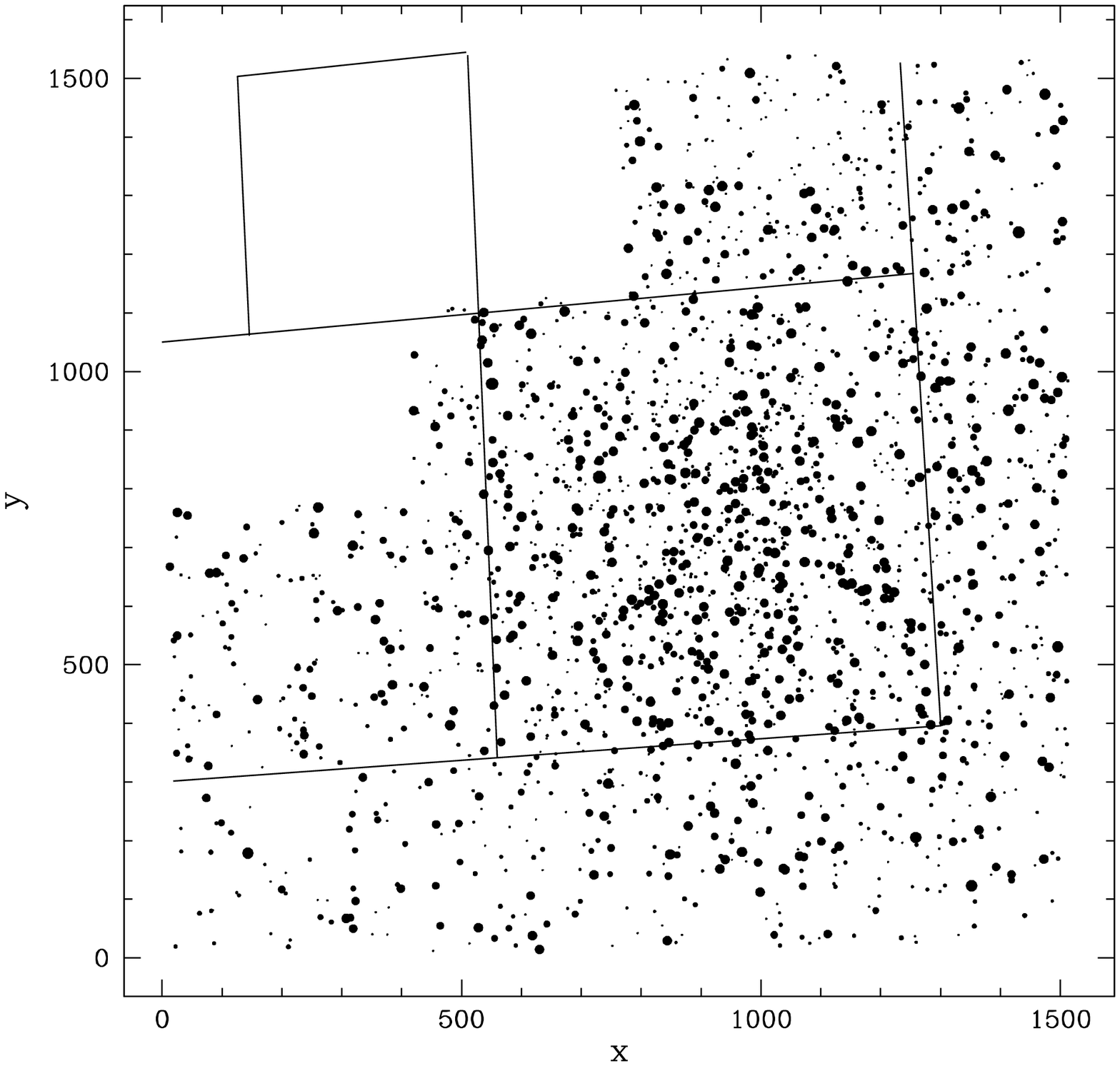,angle=0,width=17cm}
\caption[ ]{X, Y pixel map of NGC 6553: all stars 
detected in the WFPC2 mosaic in
the first epoch observations are represented by black dots; symbol
sizes are proportional to the stellar brightness.  The location of the
WFPC2 mosaic of the second epoch observations is overplotted on this
map.  \label{field}}
\end{figure*}

Since the datasets were taken for independent purposes, they cover two
fields which do not completely overlap.  Figure~\ref{field} shows the
relative orientation of the mosaics.  Black dots show all stars
detected in the WFPC2 mosaic in the first epoch observations: the size
of the symbol is proportional to the star's brightness. The cluster
center is located approximately at pixel (932,682).
Overplotted on this map is the location of the WFPC2 mosaic of
the second epoch observations, where the cluster is centered on
WF3. This configuration, coupled with the different limiting
magnitudes of the two observation sets, implies that only about 1/3 of
the stars detected in the first epoch data have a counterpart in the
second epoch photometry and could be used for astrometric purposes.

All images were de--biased, dark corrected and flatfielded through
the standard HST pipeline. Following Silbermann et al. (1996), we have
masked out the vignetted region, the saturated and bad pixels and
columns using a vignetting mask, together with the appropriate data
quality file for each frame.

The photometric reduction of each dataset was carried out using the
DAOPHOT~II/ALLFRAME package (Stetson 1987, 1994).  Preliminary
photometry was performed on each single frame in order to obtain an
approximate list of stars for each of them. The coordinates of the
stars in common were used for an accurate spatial match among the
different frames.  With the coordinate transformations, we then
obtained a single image, which is the median of all the frames,
regardless of the filter. In this way we removed all the cosmic rays
and obtained the highest signal-to-noise image for star finding. The
DAOPHOT/FIND routine was applied to the median image and Point Spread
Function (PSF) fitting photometry was carried out, in order to obtain
the deepest list of stellar objects, free from spurious
detections. Finally, this list was used as input to ALLFRAME, for the
simultaneous profile fitting photometry of individual frames. The PSFs
that we used were the WFPC2 model PSFs extracted by P. B. Stetson
(1995, private communication) from a large set of uncrowded and
unsaturated images. It is worth emphasizing that the photometric
reductions of the two datasets were carried out separately.

The quality of the first epoch photometry turned out to be better than
that of the second epoch, even if the total exposure time for each
filter was longer in the second epoch data. The difference in the
quality of the photometry in the two epochs is likely to be due to a
combination of two effects: the worsening with time of the charge
transfer efficiency loss of the WFPC2 chips, and the fact the adopted
model PSFs, being obtained in 1995, reproduced the actual PSF of the
first epoch better than the second. In what follows we will base our
analysis on the 1994 photometry, while the spatial information from the
1998 data combined to that of 1994 will be used for the relative
proper motion derivation.

Calibration of the HST instrumental magnitudes to the Landolt system
was performed following the prescriptions by Dolphin (2000). These are
an update of the widely used Holtzman et al. (1995) equations,
including the most recent correction for the charge transfer efficiency
loss. In principle, since the HST filter passbands are different from the
Landolt ones, the recommended procedure to transform the instrumental
magnitudes into the standard ground-based system is to subtract the
absolute extinction in the instrumental bands {\it before} applying
the transformations (Holtzman et al.\ 1995). However, in this
case no dereddening correction has been applied to our instrumental
magnitudes, because we were mainly interested in producing a CMD as
similar as possible to that of the ground-based photometries, and to
derive the cluster reddening. The latter, in particular, is found to
be strongly variable across the cluster area (Heitsch \& Richtler
1999, see also Sect.~5) and therefore we would anyway subtract an
inappropriate reddening to many stars.

\section{THE COLOR MAGNITUDE DIAGRAM}

\begin{figure*}
\psfig{file=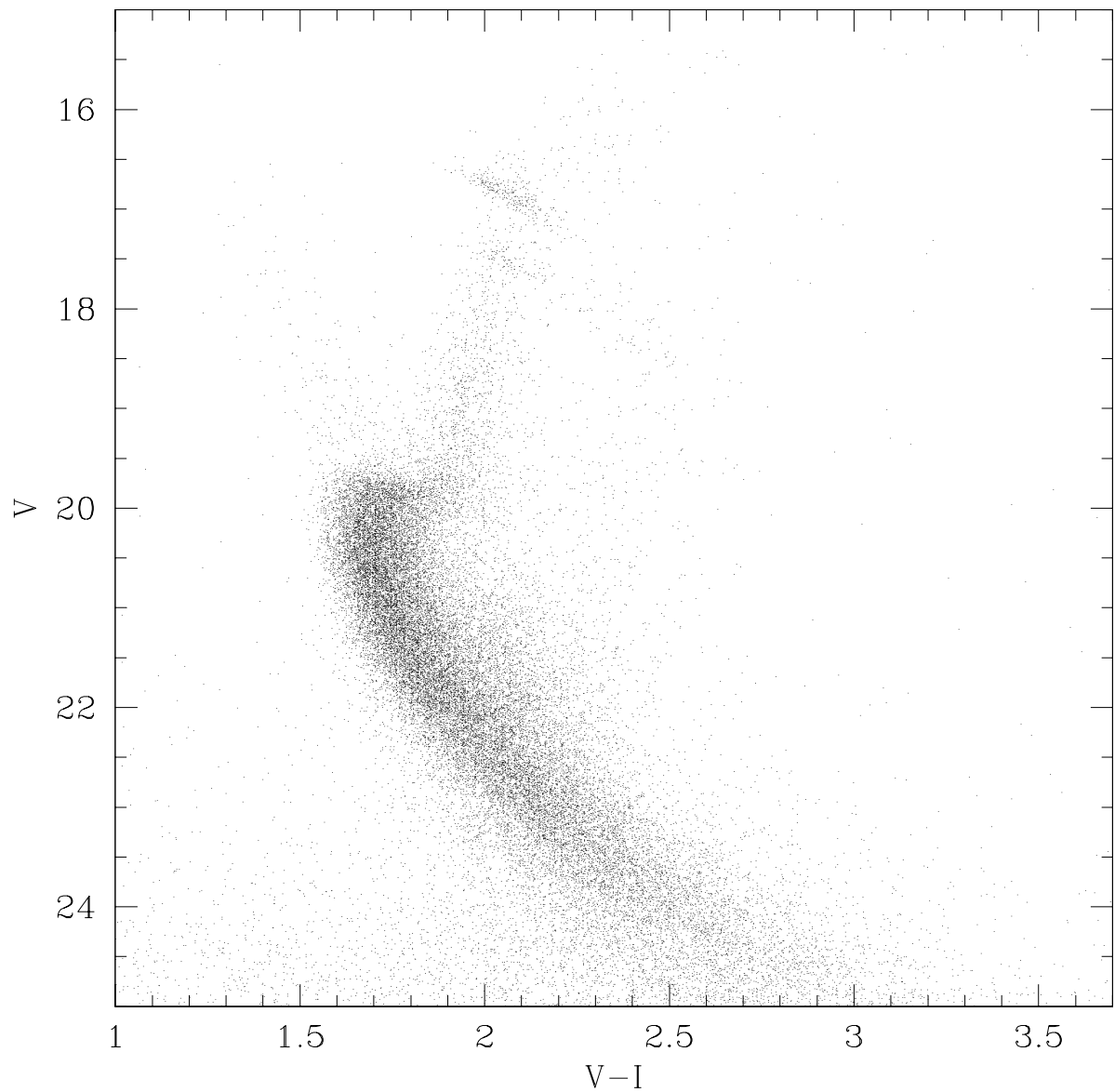,angle=0,width=17cm}
\caption{ NGC 6553 calibrated CMD extracted from the four WFPC2 chips
of the first epoch observations.\label{raw}}
\end{figure*}

Figure~\ref{raw} shows the calibrated CMD extracted from the first
epoch data. All the stars detected in the four WFPC2 chips are shown
in this diagram.  The most populated branches in this CMD are the
cluster Main Sequence (MS), Red Giant Branch (RGB), and Horizontal
Branch (HB): the latter being the inclined clump of stars at
$V\sim16.6$. The narrow sequence of stars at $V\sim17.5$, wich is
almost parallel to the HB sequence is the RGB bump (Iben, 1968); as
predicted by theory (e.g., Rood 1972), it is prominent in this
metal-rich cluster.  Both the RGB bump and the HB are shaped as an
inclined sequence (instead of a clump) by differential reddening
across the field (see below).  Also clearly visible is the bulge RGB,
which is about 0.5 magnitudes redder and $\sim 1.5$ mag fainter than the
cluster RGB (the cluster is $\sim 3$ kpc closer to us along the line
of sight than the bulk of the bulge field star population;
(see below), being the cluster $\sim3$ kpc closer to us
compared with the bulk of bulge stars.  The bulge Sub Giant Branch and
turnoff region at $V\sim21.5$ and $(V-I)\sim 2.1$ overlaps with the MS
of NGC~6553. A well populated blue sequence departing from the cluster
turnoff is also evident, and is expected to be a mixture of cluster
blue stragglers and foreground disk stars.

\begin{figure*}
\psfig{file=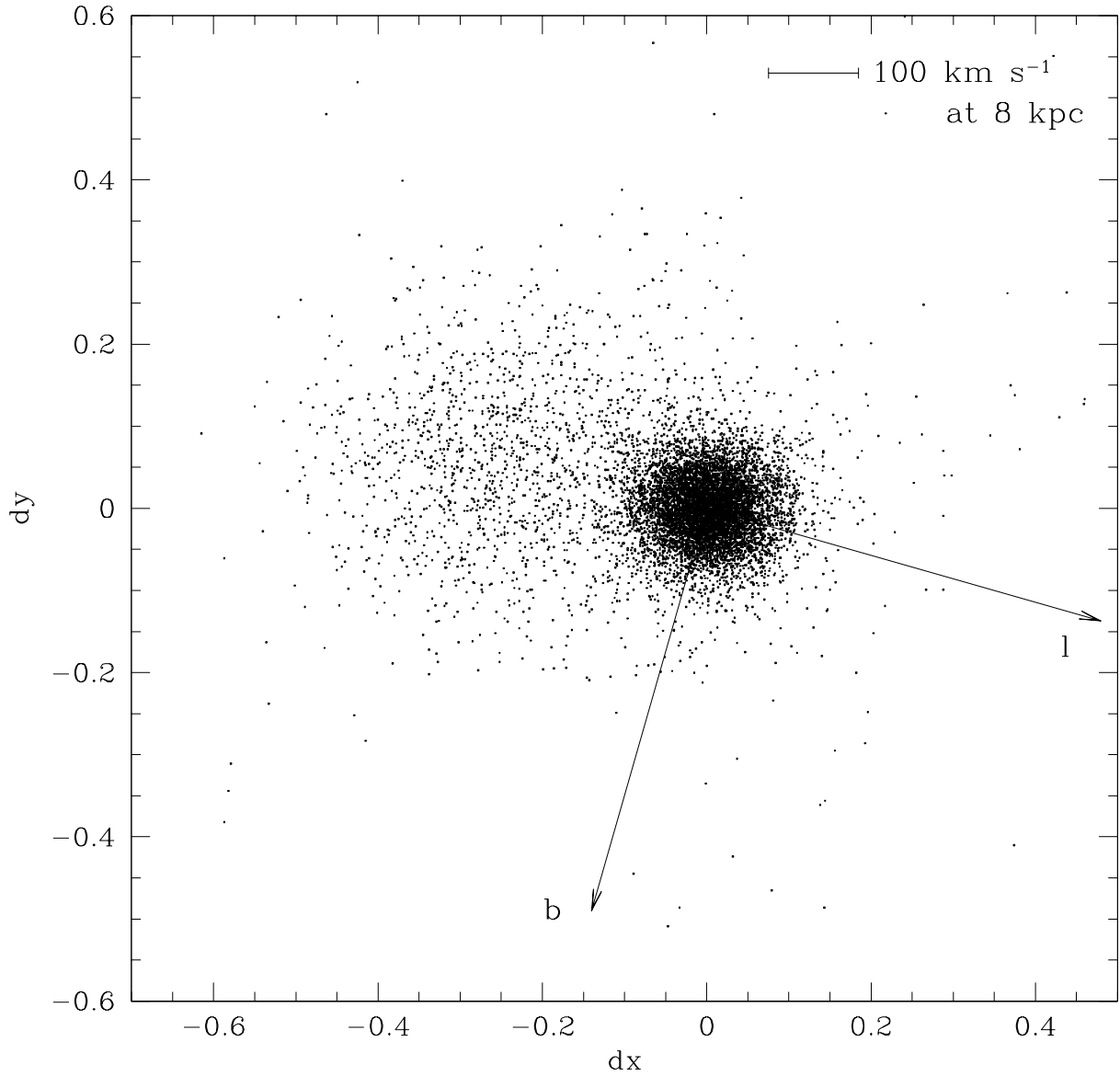,angle=0,width=17cm}
\vspace{0cm}
\caption[ ]{Relative proper motions (dx,dy) in pixels between the
two epochs. The arrows show the direction of increasing Galactic
longitude (l) and latitude (b).}
\label{dxdy}
\end{figure*}

For the astrometric calculations only those stars in common between
the two epoch data, having photometric error smaller than 0.07 mag in
both bands and shape parameter $-0.02<$SHARP$<0.05$ were
selected. Polynomial 20-coefficient coordinate transformations were
then calculated in order to take into account both the differences in
the pointings and the distortion of the WFPC2 field. Figure~\ref{dxdy}
shows the residuals differences between the coordinates of the two
epochs (epoch2-epoch1).  Obviously, since the astrometric
transformations were constrained mainly by the much more numerous
cluster stars, the residuals are, by definition, distributed around
zero.  Foreground and background (mostly bulge and some disk) stars,
instead, are distributed around $(dx,dy)=(-0.23,0.05)$ in this
plot. Also shown in Fig.~\ref{dxdy} are the directions of increasing
Galactic longitude and latitude.

\begin{figure*}
\psfig{file=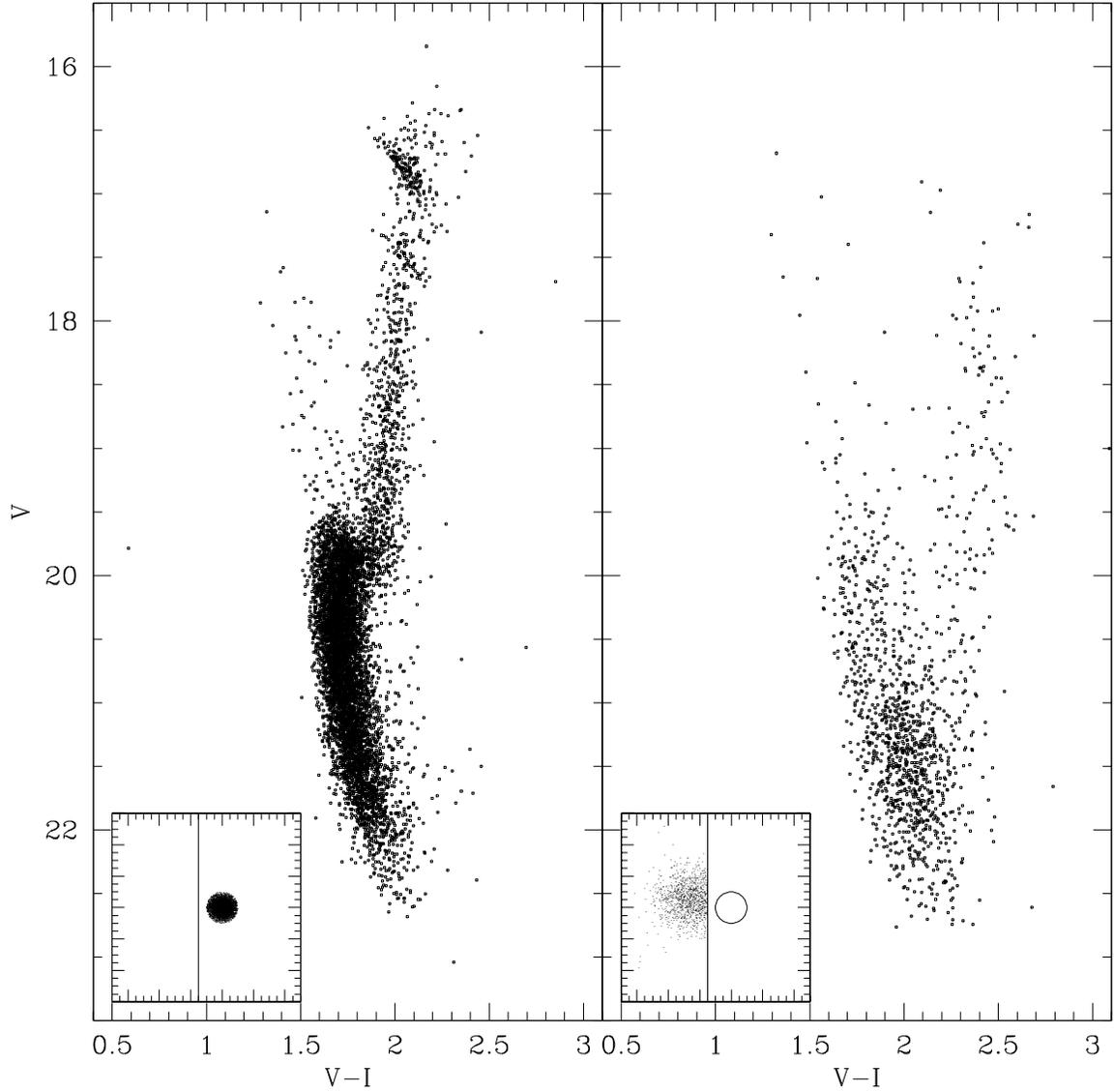,angle=0,width=17.0cm}
\vspace{0cm}
\caption[ ]{Left panel: NGC~6553 stars, selected as those lying
inside a radius of 0.1 pixels centered on (0,0) in Fig.~\ref{dxdy}
(reproduced in the panel inset).  We selected the stars inside a
relatively small radius in order to obtain the cleanest possible
cluster CMD.  Right panel: CMD of (mainly) bulge stars, selected as
those having $dx<-0.15$ in Fig.~\ref{dxdy}. Cluster stars are still
present in this CMD because, being much more numerous, a still
appreciable number of them fall well ouside the selected cluster
circle. }
\label{astrom}
\end{figure*}

Figure~\ref{astrom} shows the result of the astrometric
decontamination of the CMD. The left panel shows the CMD of only those
stars which displacement between the two epochs lies inside a radius
of 0.1 pixels centered on (0,0) in Fig.~\ref{dxdy}, which, as
expected, are mainly cluster stars. Many cluster stars fall outside
this circle, but such a small radius was chosen in order to select the
cleanest sample of cluster stars. The right panel of Fig.~\ref{astrom}
show the stars having proper motions $dx<-0.15$. Those are mainly
bulge stars, with some residual contamination from cluster outliers.

\section{PROPER MOTIONS AND SPACE VELOCITY}

\subsection{NGC~6553 Proper Motion}

The distribution of the residuals shown in Fig.~\ref{dxdy} allows us
to determine both the relative proper motion of NGC~6553 with respect
to the bulge, and the dispersion of the bulge proper motions along the
two Galactic coordinates.

The former is the difference between the centroids of the bulge and
cluster proper motion distributions, obtained by means of a Gaussian
fit along both the $l$ and $b$ axis. Figure~\ref{gauss1} shows the
histograms of the $dl$ and $db$ distributions of both cluster and
bulge stars. Note that the ratio of the integrals of the two Gaussians
in Fig.~\ref{gauss1} (lower left) also gives the relative number of
bulge to cluster stars in the region of the WFPC2 field in common
between the two epochs, which is 1400:7900=1:5.6.  The relative motion
of the cluster stars is by construction centered on $dl=0$ and $db=0$,
while the bulge stars are clustered around $(dl,db)=(-0.245,-0.017)$
pixels, which corresponds to a relative proper motion of NGC~6553 with
respect to the bulge of $\mu_l=5.89$ and $\mu_b=0.42$ mas yr$^{-1}$.
Combining this result with the radial velocity of NGC~6553, $v_r=-6.5$
(Harris 1996, as updated on the WEB page:
http://physun.physics.mcmaster.ca/Globular.html ), and adopting a Sun
motion of (U$_\odot$,V$_\odot$,W$_\odot$) = (10, 5.25, 7.17) km
s$^{-1}$ (Dehnen \& Binney 1998), the rotational velocity of the Local
Standard of Rest of $V_{\rm LSR}=239$ km s$^{-1}$ (Arp 1986) and a
cluster distance $d=5300$ pc (see below) the three components of the
NGC~6553 absolute space velocity are derived:
($\Pi$,$\Theta$,W)=(-3.5, 230, -3) km s$^{-1}$ [$\Pi$ points radially
outwards from the Galactic center, W towards the North Galactic Pole
and $\Theta$ is oriented in the direction of Galactic rotation].  This
assumes that we measure all the background bulge stars, with those in
the near side moving on average in the opposite direction with respect
to those in the far side because of bulge rotation.  Therefore, their
{\it average} proper motion is zero. One concludes that NGC~6553 is
describing a circular orbit close to the Galactic plane. Its
rotational velocity is consistent with the mean rotation of the bulge
at 2.7 kpc (Minniti 1995) but also with the disk rotation at the same
distance (e.g.\ Amaral et al.\ 1996).  Hence the kinematics of
NGC~6553 appears to be consistent with either a disk or a bulge
membership.  However, its stellar population (age and metallicity) is
identical to that of another metal rich cluster, NGC~6528 (c.f.,
Sec.~5), suggesting that the two objects have formed within the same
environment (Ortolani et al.\ 1995). NGC~6528 is located at 1.3 kpc
from the Galactic center, and its high radial velocity ($v_r=185$
km/s; Harris 1996) clearly indicates that it is on a highly eccentric
orbit, therefore excluding disk membership for this {\it bulge}
cluster. All in all, both clusters appear to belong to the bulge
population of globular clusters, with their kinematical properties
being within the corresponding distribution for bulge stars.

\subsection{Bulge Proper Motion Dispersion}

\begin{figure*}
\psfig{file=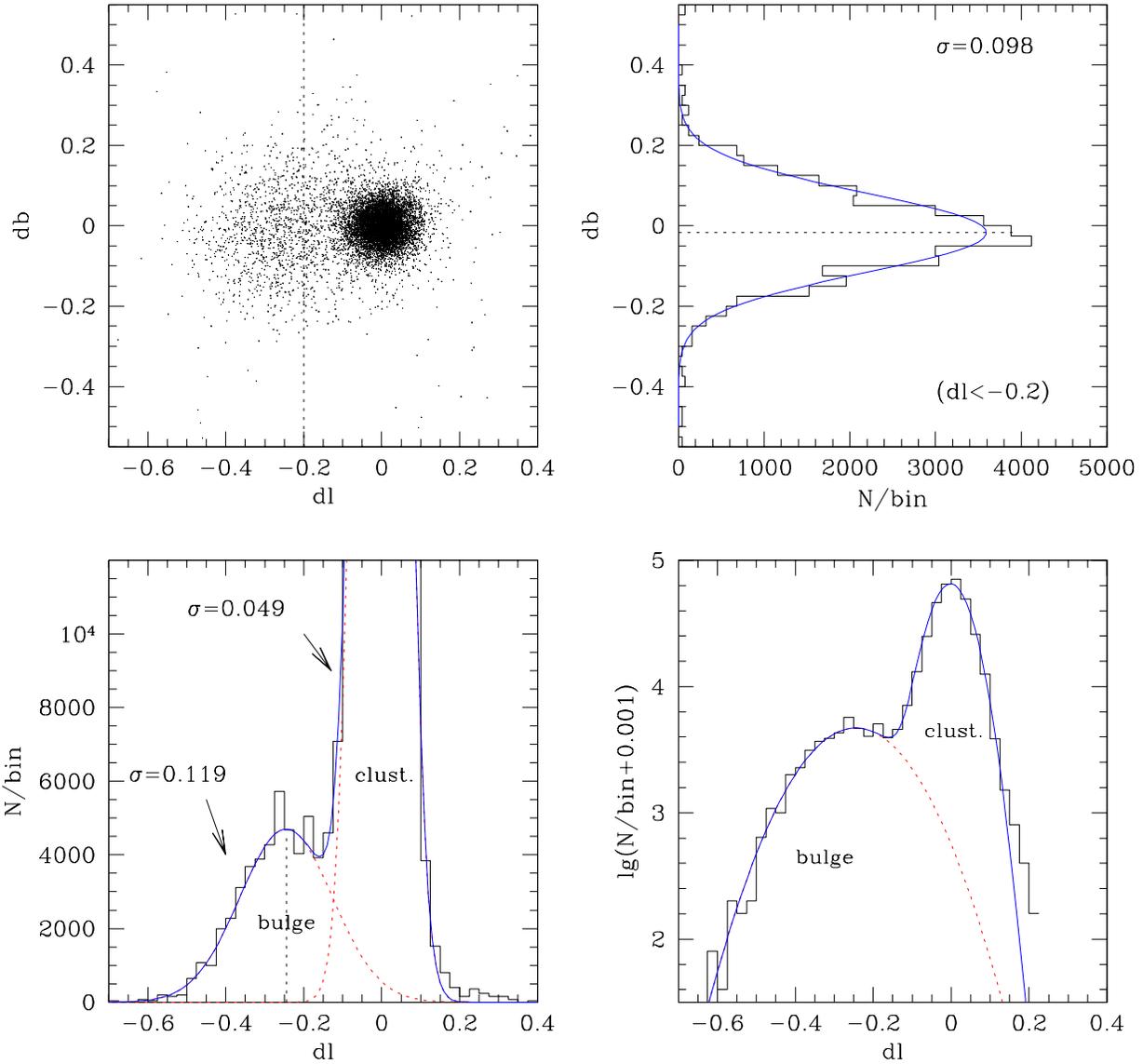,angle=0,width=17.0cm}
\vspace{0cm}
\caption[ ]{Upper left: Same as Fig.~\ref{dxdy} now as a function
of Galactic longitude and latitude. Upper right: Gaussian fit to
the $db$ distribution of stars with $dl<-0.2$ in the $(dl,db)$ plane.
Lower left: Gaussian fit to the $dl$ distribution of all the stars.
The dotted lines are the individual Gaussians while the solid line
is their sum. Lower right: same fit plotted in logaritmic units. }
\label{gauss1}
\end{figure*}

\begin{figure*}
\psfig{file=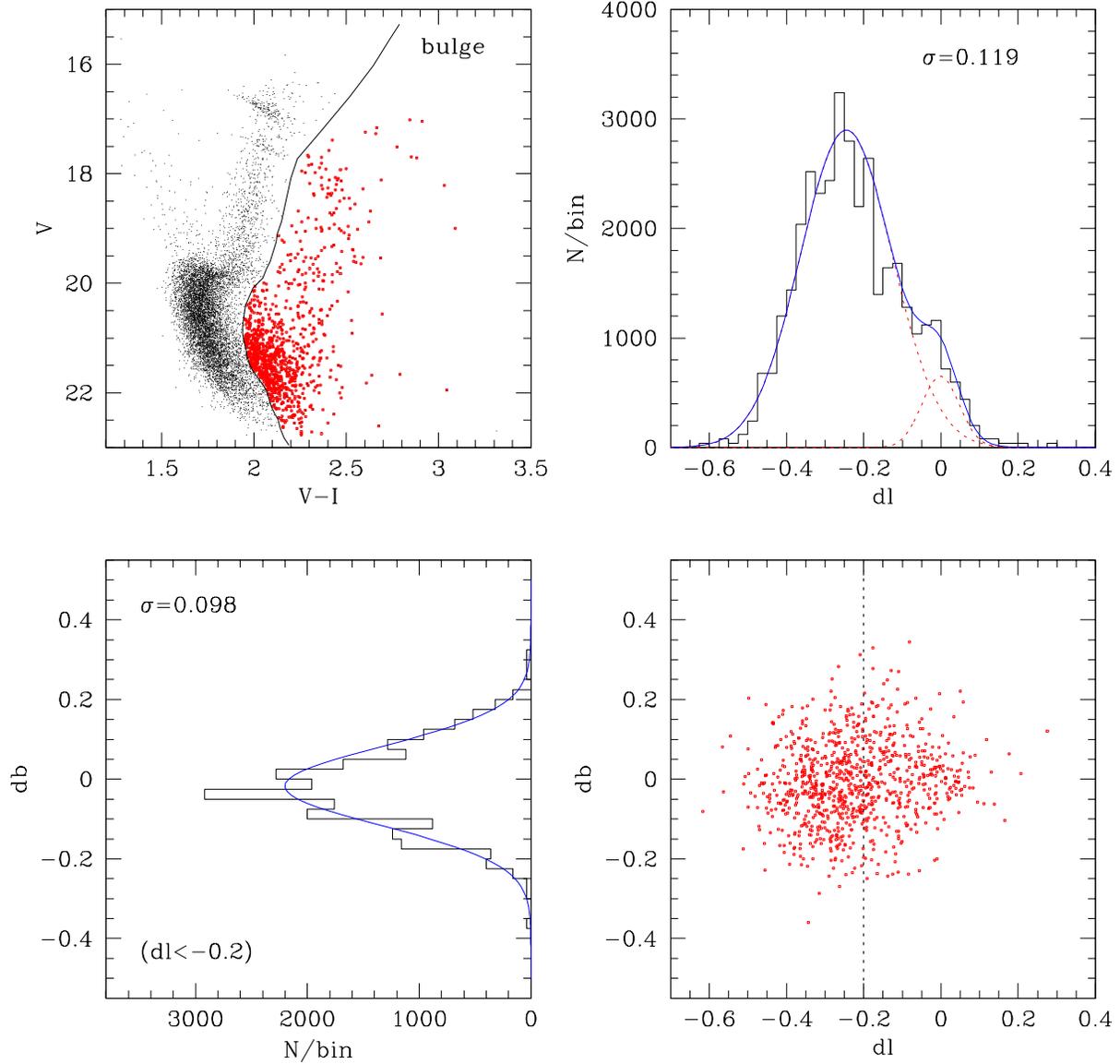,angle=0,width=17.0cm}
\vspace{0cm}
\caption[ ]{Upper left: bulge stars as selected from the CMD.
Upper right: Gaussian fit to the $dl$ distribution of all the stars
selected from the CMD. Some residual cluster members are left and were
fitted with a Gaussian with the same $\sigma$ found in
Fig.~\ref{gauss1}. The dotted lines are the individual Gaussians while
the solid line is their sum. Lower left: same fit for the $db$
distribution of stars having $dl<-0.2$. Lower right: $(dl,db)$
distribution of the bulge stars selected from the CMD. }
\label{gauss2}
\end{figure*}

The $\sigma$ of the distributions shown in Figure~\ref{gauss1} also
allow the determination of the bulge proper motion dispersion. In
order to reduce the contamination due to cluster stars in the
determination of the bulge velocity dispersion $\sigma_l$ along the
axis of Galactic longitude, bona fide bulge stars were selected from
the CMD as indicated in Fig.~\ref{gauss1}.  The distribution of these
stars in the $(dl,db)$ plane (upper right panel) gives indeed a
cleaner sample of bulge stars, with some residual cluster members,
coming from the cluster lower main sequence.  The Gaussian fit to this
distribution shown in the lower panels yields a better determination
of the bulge proper motion dispersion:
$\sigma_l=0.119\pm0.012$ and $\sigma_b=0.098\pm0.009$. The quoted
uncertainties are the formal errors in the Gaussian fit of the
histogram shown in Fig.~\ref{gauss1}. The fit was performed taking
into account errors in both coordinates: the Poisson error for the $y$
and the $\sigma$ of the cluster Gaussian for the $x$.  The dotted
lines in the lower left panel are the two Gaussians whose sum is the
solid line. The residual cluster component was fitted with a Gaussian
with the same $\sigma$ obtained from the fit of
Fig.~\ref{gauss1}. Formally, the space velocity dispersion
corresponding to the measured $\sigma$ of the cluster stars is 28 km
s$^{-1}$. We are not aware of any measurements of the velocity
dispersion of NGC~6553, but in general globular clusters have velocity
dispersions $<10$ km s$^{-1}$ (e.g., Pryor \& Meylan 1993). Therefore,
we conclude that the distribution of the cluster star in the $(dl,db)$
plane is entirely due to observational errors.  Deconvolving the
latter from the observed dispersion of the bulge stars we finally
obtain $\sigma_l=2.63\pm0.29$ and $\sigma_b=2.06\pm0.21$ mas
yr$^{-1}$.  We therefore confirm the existence of a proper motion
anisotropy $\sigma_l/\sigma_b=1.28\pm0.19$ already found by many
authors (e.g., Zhao, Rich \& Biello, 1996, who measured
$\sigma_l/\sigma_b=1.15\pm0.06$ towards Baade's Window) and
interpreted as evidence of the mean bulge rotation. The values of the
proper motion dispersions obtained from Fig.~\ref{gauss2} for our
field at $(l=5\gradip25,b=-3\gradip02)$ are somewhat lower than the
values $\sigma_l\sim3.2$ and $\sigma_b\sim 2.7$ found both towards
Baade's Window ($l=1^\circ,b=-4^\circ$) and the Plaut's Field
($l=0^\circ,b=-8^\circ$) (Spaenhauer, Jones \& Whitford 1992; Rich \&
Terndrup 1997; Mendez et al.\ 1997).

\section{THE DISTANCE AND AGE OF NGC~6553}

\begin{figure*}
\psfig{file=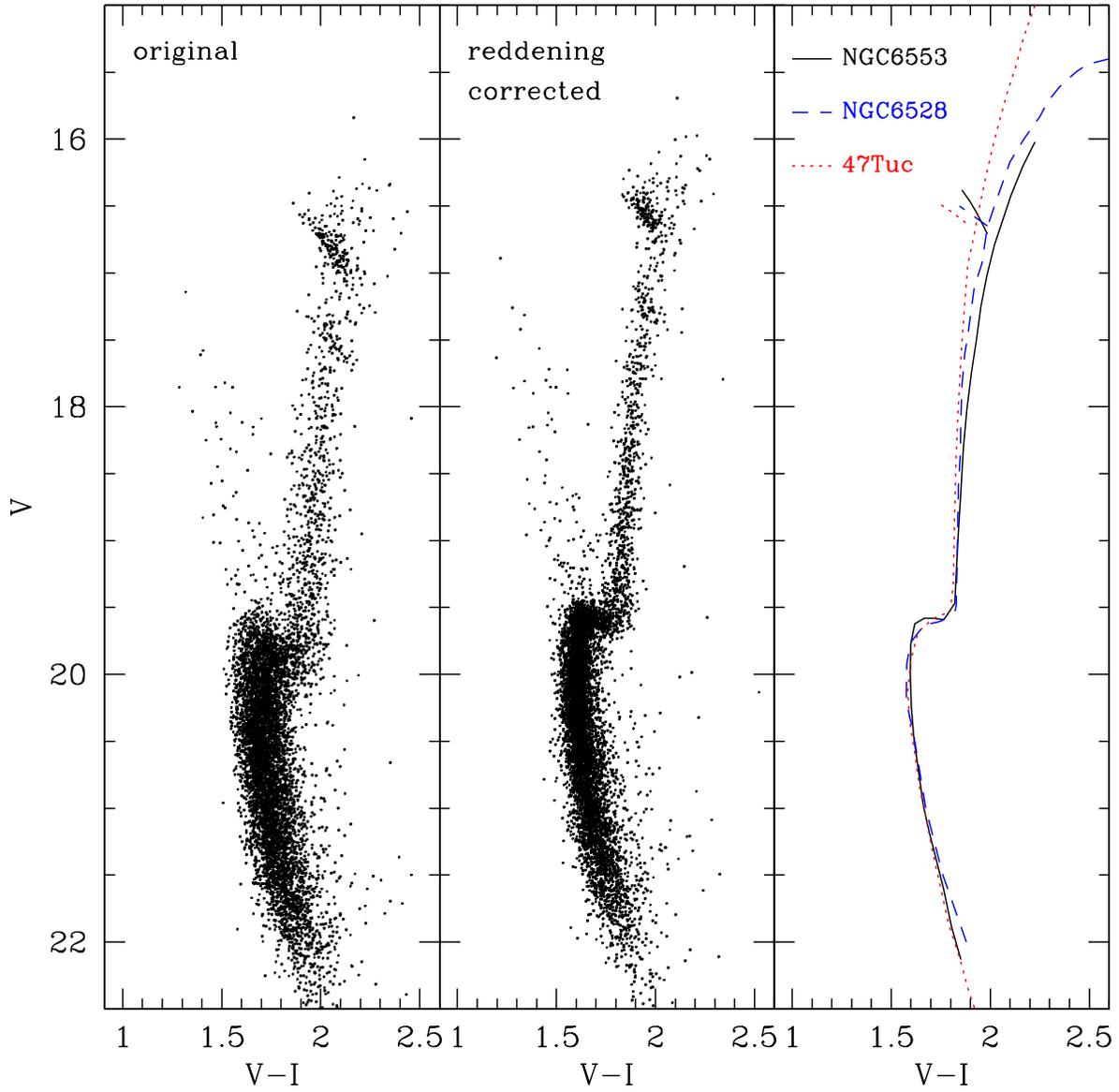,angle=0,width=17cm}
\vspace{0cm}
\caption[ ]{Left: original decontaminated CMD. Center: decontaminated
CMD also corrected from differential reddening. Right: comparison between
the fiducial lines of NGC~6553, NGC~6528 and 47~Tuc. The latter two have
been shifted both in color and magnitude in order to overlap the MS and SGB
of NGC~6553. }
\label{wf4}
\end{figure*}

The relative proper motion decontamination has provided a cleaner
cluster CMD, but the sequences still show a significant width due to
differential reddening (e.g. Heitsch \& Richtler 1999).  In the process
of matching the overlapping regions of different frames in the two
epochs data, the WFPC2 frame was divided into small sub-fields (e.g.,
part of the first epoch WF4 overlaps partly with the WF3 and partly
with WF4 of the second epoch data). In these comparisons, it appeared
that a region of the cluster mapped by the WF4 presents lower
reddening, both absolute and differential, and its CMD has much
narrower and bluer sequences. A fiducial ridge line extracted from it
was used to apply the method described in Piotto et al.\ (1999) and
obtain a cluster CMD partially corrected for differential
reddening. The observed field was divided in 84 sub-fields of equal
size and the CMD of each of them was shifted along the reddening line
to overlap the fiducial sequence. The result is show in
Fig.~\ref{wf4}: the left panel shows the original cluster CMD, the
central one is reddening-corrected and the right panel shows the 
fiducial line for 47~Tuc and NGC~6528 overplotted on that of NGC~6553. 

The central and right panels of Fig.~\ref{wf4} allow a measure of age
and distance with higher accuracy with respect to the previously
available data.  The estimated apparent magnitude of the cluster HB is
$V^{\rm HB}=16.6\pm0.05$.  Following Ortolani et al. (2000), we adopt
$M_{\rm V}^{\rm HB}=(0.16\pm0.10)\times{\rm [Fe/H]}+(0.98\pm0.10)$ 
and [Fe/H]=$-0.5\pm0.3$, giving $M_{\rm V}^{\rm HB}= 0.90\pm0.12$.
The cluster apparent distance modulus is therefore
$(m-M)_V=16.60-0.90=15.70\pm0.13$.

The reddening is derived from comparisons to the 47~Tucanae locus
(Ortolani et al.\ 1995; Kaluzny et al. 1997).  The color of the RGB of
NGC~6553 at the HB level is $V-I$ = 2.08$\pm$0.03. With respect to 47
Tuc we derive $\Delta(V-I)_{\rm (NGC6553-47Tuc)}$ = 0.99.  Part of
this color shift ($\Delta(V-I)=0.134$, Girardi et al. 2000) is due to
the metallicity difference between the two clusters.  Indeed, 47~Tuc
has metallicity [Fe/H]=--0.76 (Harris 1996) and [$\alpha$/Fe]=0.3
(Gratton, Quarta \& Ortolani 1986), therefore a global metallicity
[M/H]=--0.55, while Coelho et al. (2000) measured [M/H]=-0.1 for
NGC~6553.  Assuming $E(V-I)_{\rm 47Tuc}=0.05$ (Barbuy et al. 1998) we
finally obtain $E(V-I)_{\rm NGC6553} = 0.99 - 0.135 - 0.05 = 0.805$
and therefore $E(B-V)=E(V-I)/1.28=0.63$ and $A_{\rm
V}=0.63\times3.3=2.08$.  The adopted total-to-selective extinction
ratio $R_{\rm V} = 3.3$ ratio was assumed considering the metallicity
and reddening amount dependences (Barbuy et al. 1998 and references
therein). This value of the reddening is lower than previous
determinations (and therefore the {\it apparent} distance modulus
shorter) because the reddening corrected CMD has been registered on
the fiducial line of a low-reddening region.  The absolute distance
modulus of NGC~6553 is $(m-M)_0=15.70-2.08=13.62$, in perfect
agreement with the value $(m-M)_0=13.60$ found by Guarnieri et al.\
(1997), and corresponding to a heliocentric distance $d=5.3$ kpc.

Figure~\ref{wf4} also shows the fiducial line of 47~Tuc and NGC~6528
(Ortolani et al. 1995) overplotted on that of NGC~6553. The former two
were shifted both in magnitude and in color in order to bring their
upper MS into coincidence with that of NGC~6553.  The extremely small
differences in the TO and HB locations are well inside the uncertainty
in the construction of the fiducial lines, from which we conclude that
the three clusters are coeval to within $\sim 1$ Gyr, and therefore
the absolute age of NGC~6553 (and NGC~6528) is $\sim13$ Gyr as
recently determined for 47~Tuc (Zoccali et al.\ 2001). This
demonstrates that the Galactic bulge was enriched to solar abundance
to within a rather short time scale (less than a few Gyr). The
difference in the RGB slopes seen in the right panel of Fig.~\ref{wf4}
is due to the difference in global metallicity of the three clusters.
Note that the age-sensitive parameter $\Delta$V$^{\rm TO}_{\rm HB}$
changes by only a few hudredths of magnitudes in this metallicity
range (see Fig.~3 of Rosenberg et al. 1999).

\section{THE CLUSTER BLUE STRAGGLERS}

As apparent from Fig.~\ref{astrom}, at least part of the stars in the
blue sequence above the cluster MS seem to belong to the cluster,
hence they are probably blue stragglers (BS), rather than Galactic
disk main sequence stars. This is confirmed by the following two
experiments.

\begin{figure*}
\psfig{file=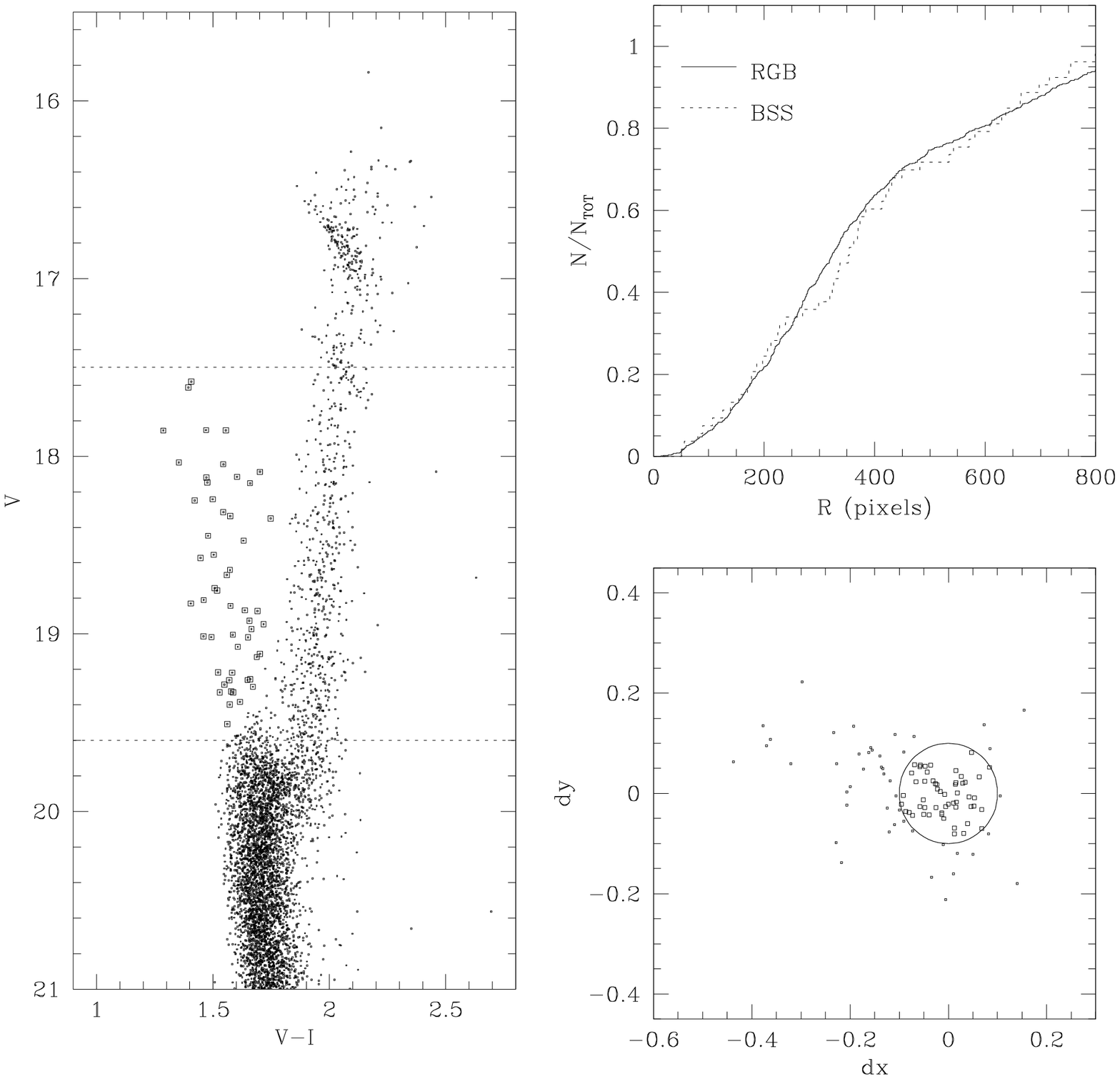,angle=0,width=17.0cm}
\vspace{0cm}
\caption[ ]{Left panel: The open squares indicate the candidate BS in the
CMD of the likely cluster members, i.e., those stars with proper
motion $<0.1$ pixels.  Upper right panel: cumulative radial
distribution (in pixels) of the BS (dotted line) compared with the RGB
stars (solid line) with $17.5<V<19.6$.  Lower right: proper motion
distribution of all the stars brighter than $V=19.3$ and bluer than
$V-I=1.76$ in the original CMD. Note the strong concentration within
the circle of 0.1 pixels radius, chosen to select the most likely
cluster members. }
\label{bss}
\end{figure*}

In order to check this issue, we isolated the probable BS stars in the
decontaminated CMD (Fig.~\ref{bss}: left panel), and investigated
their radial distribution as compared with the other cluster
stars. The upper right panel of Fig.~\ref{bss} shows that the
cumulative distribution of the BS (dotted line) have the same spatial
distribution as the cluster RGB (solid line) with $17.5<V<19.6$. This
indicates that they are indeed cluster members.  However, the BS
population of NGC~6553 does not appear concentrated towards the center
as they are in high density clusters (e.g., Ferraro et al. 1999). The
second check on the BS nature of a major fraction of the blue sequence
stars can be made by using their distribution of proper motions. This
is displayed in the lower right panel of Fig.~\ref{bss} which shows
{\it all} the stars brighter than $V=19.3$ and bluer than $V-I=1.76$
in the original CMD (Fig.~\ref{raw}), and confirms that a large
fraction of them are indeed concentrated around $(dx,dy)=(0,0)$ and
therefore are likely BS members of NGC~6553.

\section{CONCLUSIONS}

Using WFPC2 images taken 4.16 yr apart we have been able to measure
both the proper motion of the globular cluster NGC 6553 and of the
Galactic bulge stars in the background of the cluster. The main
results obtained can be summarized as follows.

\pn
1. The cluster NGC 6553 appears to be on a nearly circular orbit $\sim
3$ kpc from the center of the Galaxy, with a velocity of $\sim 230$ km
s$^{-1}$ and a small inclination with respect to the Galactic plane.
These kinematical properties are consistent with both a bulge and a
disk membership of the cluster. However, the cluster stellar
population is virtually identical to that of the cluster NGC 6528,
which kinematics makes its membership to the bulge unambiguous, and we
conclude that also NGC 6553 is likely a member of the Galactic bulge.

\pn
2. The relative astrometry between the two epochs is accurate enough
for measuring the dispersion of proper motions of bulge stars along
the two Galactic coordinates, with $\sigma_l=2.63$ and $\sigma_b=2.06$
mas yr$^{-1}$.
\pn
3. The distinct proper motions of the cluster and bulge stars has
allowed us to construct decontaminated CMDs for both the cluster and
bulge populations. The resulting improved CMD for the cluster has
allowed us to obtain more accurate values for the reddening, distance,
and age of the cluster: NGC 6553 appears to have the same age as the
inner halo globular cluster 47 Tuc ($\sim 13$ Gyr) within a $\sim 1$
Gyr uncertainty. Given the high metallicity of both NGC 6553 and NGC
6528, this indicates that the Galactic bulge underwent rapid metal
enrichment to solar metallicity and beyond, some 13 Gyr ago,
supporting the view of an early and rapid assembly of the Galactic
Bulge.

\pn
4. Thanks to the accurate astrometry obtained, we have also
demonstrated that the cluster contains a significant population of
blue stragglers, having separated them from a trace population of
stars with similar luminosities and colors which instead are likely
intermediate age foreground disk stars.


\acknowledgements
We thank the referee, Patrick C\^ot\'e, for useful comments.
BB and EB acknowledge partial financial support from CNPq and Fapesp
(Brasil). SO acknowledges the Ministero dell'Universit\`a e della
Ricerca Scientifica e Tecnologica (MURST) under the program on
'Stellar Evolution' (Italy).

%

\newpage

\newpage

\newpage

\newpage

\newpage

\newpage

\newpage

\begin{references}

\reference{} Amaral, L.H., Ortiz, R., L\'epine, J.R.D.,
        Maciel, W.J. 1996, MNRAS, 281, 339
\reference{} Armandroff, T.E. 1989, AJ 97, 375
\reference{} Arp, H. 1986, A\&A, 156, 207
\reference{} Barbuy B., Bica E., Ortolani S. 1998, A\&A, 333, 117
\reference{} Barbuy, B., Renzini, A., Ortolani, S., Bica, E., \& Guarnieri, 
	M.D. 1999a, A\&A, 341, 539
\reference{} Barbuy B., Ortolani S., Bica E., Desidera S. 
	1999b, A\&A 348, 783
\reference{} Coelho, P., Barbuy, B., Perrin, M.N., Idiart, T.,
        \& Schiavon, R.P., 2000, A\&A, in preparation
\reference{} C\^ot\'e, P., 1999, AJ, 118, 406
\reference{} Dehnen, W., \& Binney, J.J. 1998, MNRAS, 298, 387
\reference{} Djorgovski, S. 1993, in Structure and Dynamics of Globular
	Clusters, eds.\ S.\ G.\ Djorgovski \& G.\ Meylan (San Francisco:\ ASP),
	p.\ 373
\reference{} Dolphin A.E., 2000, PASP, 112, 1397
\reference{} Ferraro, F.R., Paltrinieri, B., Rood, R.T., Dorman, B., 
	1999, ApJ, 522, 983
\reference{} Girardi, L., Bressan, A., Bertelli, G., \& Chiosi, C.
        2000, ApJ, 530, 62
\reference{} Gratton, R.G., Quarta M.L., \& Ortolani S. 1986,
        A\&A, 169, 208
\reference{} Guarnieri, M.D., Renzini, A., \& Ortolani, S.
        1997, ApJ, 477, 21
\reference{} Guarnieri M.D., Ortolani S., Montegriffo P.,
        Renzini A., Barbuy B., Bica E., Moneti A., 1998, A\&A 331, 70
\reference{} Harris W.E. 1996, AJ 112, 1487
\reference{} Harris, W.E. 2000, Saas Fee Lectures 1998 (Berlin, Springer), 
	in press
\reference{} Heitsch, F., \& Richtler, T. 1999, A\&A, 347, 455
\reference{} Holtzman J., Burrows C.J., Casertano S. et al.,
        1995, PASP 107, 1065
\reference{} Iben, I. Jr. 1968, ApJ, 154, 581
\reference{} Kaluzny, J., Krzeminski, W., Mazur, B., Wysocka, A.,
        Stepien, K. 1997, Acta Astron., 47, 249
\reference{} King I.R., Anderson J., Cool A.M., Piotto G., 1998, ApJ,
        492, L37
\reference{} Mendez, R.A., Rich, R.M., Van Altena, W.F.,
        Girard, T.M., van den Berg, S., \& Majewski, S., 1998, IAU Symp.
\reference{} Minniti, D., 1995, AJ, 109, 1663
\reference{} Ortolani S., Barbuy B., Bica E.
        1990, A\&A 236, 362
\reference{} Ortolani S., Renzini A., Gilmozzi R., Marconi G.,
        Barbuy B., Bica E.,  Rich R.M. 1995, Nature 377, 701
\reference{} Ortolani, S., Momany, Y., Bica, E., \& Barbuy, B.,
        2000, A\&A, 357, 495
\reference{} Piotto, G., Zoccali M., King I.R., Djorgovski S.G.,
        Sosin C., Rich R.M., Meylan G., 1999, AJ, 118, 1737
\reference{} Pryor, C., \&, Meylan, G., 1993, in ASP Conf.Ser.50, Structure
        and Dynamics of Globular Clusters, eds. S.G. Djorgovski and G. Meylan,
        (San Francisco: ASP), 357
\reference{} Rich, R.M., \& Terndrup, D.M., 1997, in ASP Conf.Ser. 127,
        Proper Motions and Galactic Astronomy, ed. R.M. Humphreys, (San
        Francisco: ASP), 129
\reference{} Rood, R.T., 1972, ApJ, 177, 681
\reference{} Rosenberg, A., Saviane, I., Piotto, G., \& Aparicio, A., 
	1999, AJ, 118, 2306
\reference{} Sagar, R., Subramaniam, A., Richtler, R., and Grebel, E.K. 1999, 
	A\&A, 135, 391
\reference{} Silbermann N.A., Harding P., Madore B.F., 1996, ApJ 470, 1
\reference{} Spaenhauer, A., Jones, B.F., \& Whitford, A.E., 1992, AJ, 103, 297
\reference{} Stetson P.B., 1987, PASP 99, 191
\reference{} Stetson P.B, 1994, PASP 106, 250
\reference{} Zhao, H., Rich, R.M., \& Biello, J., 1996, ApJ, 470, 506
\reference{} Zinn, R., 1985, ApJ, 293, 424
\reference{} Zoccali, M., Renzini, A., Ortolani, S., Bragaglia,
        A., Bohlin, R., Carretta, E., Ferraro, F.R., Gilmozzi, R., Holberg,
        J.B., Marconi, G., Rich, R.M., Wesemael, F., 2001, ApJ, submitted

\end{references}
\end{document}